\documentstyle[12pt]{article}

\newcommand{\be}{\begin{equation}}
\newcommand{\ee}{\end{equation}}
\newcommand{\ba}{\begin{eqnarray}}
\newcommand{\ea}{\end{eqnarray}}
\newcommand{\la}{\lambda}
\newcommand{\al}{\alpha}

\newcommand{\tr}{\rm tr}

\begin{document}
\hoffset=-.4truein\voffset=-0.5truein
\setlength{\textheight}{8.5 in}
\begin{titlepage}
\begin{center}
\hfill {LPTENS }\\
\vskip 0.6 in
{\large \bf Random supermatrices with an external source}
\vskip .6 in
\begin{center}
{\bf E. Br\'ezin$^{a)}$}{\it and} {\bf S. Hikami$^{b)}$}
\end{center}
\vskip 5mm
\begin{center}
{$^{a)}$ Laboratoire de Physique
Th\'eorique, Ecole Normale Sup\'erieure}\\ {24 rue Lhomond 75231, Paris
Cedex
05, France. e-mail: brezin@lpt.ens.fr{\footnote{\it
Unit\'e Mixte de Recherche 8549 du Centre National de la
Recherche Scientifique et de l'\'Ecole Normale Sup\'erieure.
} }}\\
{$^{b)}$ Okinawa Institute of Science and Technology Graduate University, 1919-1 Tancha, Okinawa 904-0495, Japan.
e-mail: hikami@oist.jp
} \\
\end{center}     
\vskip 3mm         

{\bf Abstract}                  
\end{center}
 In the past we have considered   Gaussian random matrix ensembles in the presence of an external matrix source. The reason was that it allowed, through an appropriate tuning of the eigenvalues of the source,  to obtain results on non-trivial dual models, such as Kontsevich's Airy matrix models and generalizations. The techniques relied on explicit computations of the k-point functions for arbitrary N (the size of the matrices) and on an N-k duality.  Numerous results on the intersection numbers of the moduli space of curves were obtained by this technique.  In order to generalize these results to include surfaces with boundaries,  we have extended these techniques to supermatrices. Again we have obtained quite remarkable explicit expressions for the k-point functions, as well as  a duality.  Although supermatrix models  a priori lead to the same matrix models of  2d-gravity, the external source extensions considered in this article lead to new geometric  results.  
 \end{titlepage}
\vskip 3mm

%*******************************
\section{Introduction}
 We have considered at length in the past  Hermitian random matrices in the presence of an external matrix source \cite{BH1,BH2}. In fact we have limited ourselves to Gaussian models  because a  specific duality of these models, to be recalled below,  made it possible to use the matrix source in order to tune non-trivial models such as Kontsevich's Airy matrix model\cite{Kontsevich} and generalizations \cite{Witten1}.  Such models have led to easy calculations of intersection numbers for the moduli space of  curves with marked points and boundaries \cite{BH1,BH2,DijkgraafWitten}.
 
 The triangulation of surfaces  through supermatrices should be useful to characterize super-Riemann surfaces (SRS) or super-Teichmuller space \cite{Witten2, Penner}. As  a first step to investigate the moduli space for SRS  through
 supermatrices with an external source, we compute explicitly the expectation values of the supervertices.

 We consider here a Gaussian  ensemble of supermatrices,  a generalized GUE,   in the presence of an external matrix source.  It presents a number of similarities with the usual case : (1) the k-point function $<s\tr e^{t_1M}\cdots s\tr e^{t_k M}>$ are explicitly calculable for random matrices $M$  invariant under the super-unitary group $U(n\vert m)$ or $UOSp(n\vert m)$, ( 2) there is again a dual representation of $<\prod_1^k {\rm sdet}^{-1}(x_i-M)>$  valid for arbitrary $(n,m)$ in terms of integrals over matrices of size  $k\times k$.

%%%%%%%%%%%%%%%%%%%%%%%%%%%%%%%%%%%%%%%%%%%%%%%%%%%%%%%%%%%%%%%%%%%%%
\section{One point function}

The "probability" distribution for  super-Hermitian matrices is 
\be\label{PA}
P_A(M) = \frac{1}{Z_A} e^{\frac{i}{2}{s\tr} M^2 +i  {s\tr} M A}
\ee
in which  the matrix  
\be M = \left(\begin{array}{cc}
a & \alpha \\\bar{\alpha}  & b
\end{array}\right)\ee
We have to deal  with a complex weight to make meaning of the integrals since 
\be s\tr M^2 = \tr (a^2) -\tr (b^2)+ 2 \tr {(\al {\overline \al} )}\ee
 The $n\times n$ matrix $ a$ is Hermitian , and the  
 $m\times m$ matrix $ b$ is also Hermitian ; the matrices $\al$ and $\bar\al$ are rectangular, respectively $n\times m$ and $m\times n$ and consist of Grassmanian (i.e. anticommuting) variables. We use the convention ${\overline{\al \beta}} =  {\overline{\beta}} {\overline{\al}}$.   We denote the eigenvalues of the source super-matrix $A$  by $(r_i, \rho_j)$  which we can take as a diagonal matrix.

We would like to compute the one-point function $< {\rm str} \ e^{itM} >$, 
expectation value with respect to the weight (\ref{PA}).  If we assume that $M$ may be diagonalized through a super-unitary transformation $U(n\vert m)$  , i.e. $M = U^{\dagger} D U $  with
\be D=  \left(\begin{array}{cc}
l & 0\\0  & \mu 
\end{array}\right)\ee
we can replace the integral over $M$ by an integral over its eigenvalues $l 's $ and $\mu 's$ plus an integral over the super-unitary group. (For instance if the matrix $M$ is just two by two, $l = a+ \frac{\al\bar\al}{a-b}$  and $\mu= b+ \frac{\al\bar\al}{a-b}$).  \\The usual Vandermonde Jacobian associated with this diagonalization  is replaced by the Berezinian\cite{Berezin}
 \be J(l, \mu) = (\frac { \Delta(l) \Delta(\mu)}{ \Delta (l\vert \mu)})^2 \ee
with
\be  \Delta (l\vert \mu) = \prod_{a=1}^n \prod_{b=1}^m( l_a-\mu_b) \ee
Since the observable str$e^{itM}$  is unitary invariant,  the integral over the unitary group involves only  the Itzykson-Zuber like integral
\be I = \int dU e^{is\tr
 U^{\dagger} D U A} \ee
This integral has been computed by Alfaro and co-workers \cite{Alfaro} who found
\be\label{IZ} I =\frac{ \det e^{il_i r_j} \det e^{-i\mu_i \rho_j}  \Delta(l\vert\mu) \Delta( r\vert\rho)} {\Delta(l)\Delta(\mu) \Delta(r)\Delta(\rho)} \ee
up to a normalization which will be fixed later ; the $\Delta's$ are Vandermonde factors as usual. 

Inserted  into the expression for $U(t)$ the $n!$ terms of the expansion of $\det e^{il_i r_j}$ and the $m!$ terms of $\det e^{-i\mu_i \rho_j} $ are all equal thanks of the antisymmetry of $\Delta(l)$ and $\Delta(\mu)$. 
Therefore combining the Berezinian and the IZ integral we obtain 
\ba \label{U} U(t) =&& \frac{\Delta(r\vert\rho)}{Z_A\Delta(r)\Delta(\rho)} \int dl_i d\mu_j \frac{ \Delta(l) \Delta(\mu)}{\Delta(l\vert \mu)} e^{\frac{i}{2}(\sum l_i^2 - \sum \mu_j^2) + i\sum l_ir_i-i\sum\mu_j\rho_j}\nonumber\\ &&  ( \sum_a e^{itl_a}-\sum_j e^{it\mu_j} )\ea
We now use an identity, similar to the one which we have used in the past for the usual GUE,  namely
\ba &&\label{identity} \int dl_i d\mu_j \frac{ \Delta(l) \Delta(\mu)}{\Delta(l\vert \mu)} e^{\frac{i}{2}(\sum l_i^2 - \sum \mu_j^2) + i\sum l_ar_a-i\sum\mu_j\rho_j} \nonumber\\
&&= e^{-\frac{i}{2} (\sum r_i^2-\sum\rho_j^2) }\frac{ \Delta(r)\Delta(\rho)} {\Delta(r\vert\rho)} \ea
which follows trivially from the fact that the partition function $Z_A$ in (\ref{PA}) is simply equal to $ e^{-\frac{i}{2} s\tr A ^2} $. The identity (\ref{identity}) follows from a calculation of $Z_A$ based on the diagonalization of $M$ and of the susy IZ formula (\ref{IZ}). 

In order to complete the calculation we note that each of the $(n+m)$ terms generated from the second line of (\ref{U}) involves a simple modification of the source matrix $A$. For instance the first one involves the replacement
$r_i\to r_i +t \delta_{i1}$ and since we know the integral for arbitrary $r_i's$ from (\ref{identity}) we can perform all the integrals over the eigenvalues and end up with a sum of $(n+m)$ terms. It turns out that, as in the simple GUE case, the sum of the $n$ terms as well as the sum over the $m$ terms may be replaced by one single contour integral encircling respectively  the poles at $z=r's $ and  at $z= \rho's$ . We end  up with
\be U(t) =U^I(t) + U^{II}(t) \ee
\be \label{1point}U^I(t) = \frac{e^{-it^2/2}}{t}  \oint \frac{dz}{2i\pi}e^{-itz} \frac{ \prod_{i=1}^n (1+\frac{t}{z-r_i})}{ \prod_{j=1}^m ( 1+\frac{t}{z-\rho_j})} \ee
\be  U^{II}(t) = \frac{e^{it^2/2}}{t} \oint \frac{dz}{2i\pi} e^{-itz}\frac{ \prod_{j=1}^m (1-\frac{t}{z-\rho_j} )}{ \prod_{i=1}^n ( 1-\frac{t}{z-r_i})}\ee
In the first  integral the contour  encircles the poles $z=r_i$'s and not $z=\rho_j$. Each pole provides one of the first $n$ terms of (\ref{U}). Similarly the second contour encircles the poles at $z=\rho_j$ and provides the remaining $m$ terms. 
In the course of the calculation we have dropped a number of constants since they cancelled with the normalization $Z_A$ . One can check that the final normalization is  right  since it verifies
\be U(0) = <s\tr 1> = n-m \ee
Remarkably enough if we shift $z$ to $z-t/2$ in the first integral and $z$ to $z+t/2$ in the second, one finds that $U^I(t)$ and $U^{II}(t)$ recombine into the single integral

\be U(t)\label{one} = \frac{1}{t}  \oint \frac{dz}{2i\pi}e^{-itz}  \prod_{i =1}^n \frac{z-r_i +t/2}{z-r_i-t/2} \prod_{j=1}^m \frac{z-\rho_j -t/2}{z-\rho_j+t/2} 
 \ee
in which the contour circle over all the poles  at $z= r_i+t/2$ and $z=\rho_j -t/2$. \\
In the absence of any source, i.e. if all the $r$'s and $\rho$'s vanish, the result is 
\be U(t) = \frac{1}{t}  \oint \frac{dz}{2i\pi}e^{-itz} ( \frac{z +t/2}{z-t/2})^{n-m} \ee
i.e., a simple dimensional reduction $n\to n-m$ of the GUE result \cite{BH1}, but in general it is indeed genuinely different. 

%%%%%%%%%%%%%%%%%%%%%%%%%%%%%
\section{Two point correlation function}
The same technique allows one to compute correlation functions such 
\be U(t_1,t_2) = < s\tr e^{it_1M}  s\tr e^{it_2M} > \ee
After integration over the unitary degrees of freedom one is left with
\ba  U(t_1,t_2)  =&& \frac{\Delta(r\vert\rho)}{\Delta(r)\Delta(\rho)} \int dl_i d\mu_j \frac{ \Delta(l) \Delta(\mu)}{\Delta(l\vert \mu)}) e^{\frac{i}{2}(\sum l_i^2 - \sum \mu_j^2) + i\sum l_ir_i-i\sum\mu_j\rho_j} \nonumber \\&&  ( \sum_a e^{it_1l_a}-\sum_j e^{it_1\mu_j} ) ( \sum_a e^{it_2l_a}-\sum_j e^{it_2\mu_j} )\ea
i.e. $(n+m)^2$ terms which can all be computed with the help of the identity (\ref{identity}) through an appropriate shift of the eigenvalues of the source matrix such as 
\be r_a\to r_a+t_1 \delta_{ai} + t_2 \delta_{aj} \ee
and  similarly for the $(r,\rho)$ and $(\rho, \rho)$ terms. 
This leads to a sum of four terms 

\ba &&U^I (t_1,t_2) = \frac{e^{-it_1^2/2 -it_2^2/2}}{t_1t_2}   \oint \frac{dz_1}{2i\pi} \oint \frac{dz_2}{2i\pi}e^{-it_1z_1-it_2z_2}\\ \nonumber && \frac{ \prod_{i=1}^n (1+\frac{t_1}{z_1-r_i}) (1+\frac{t_2}{z_2-r_i})}{ \prod_{j=1}^m ( 1+\frac{t_1}{z_1-\rho_j})( 1+\frac{t_2} {z_2-\rho_j})} [1+ \frac{t_1t_2} {(z_1-z_2+t_1) (z_1-z_2-t_2)}]\ea
in which both contours encircle the poles $r_i$. 
Similarly there  are three more terms; the plus-minus combination gives
\ba &&U^{II} (t_1,t_2) = \frac{e^{-it_1^2/2 + it_2^2/2}}{t_1t_2} \oint \frac{dz_1}{2i\pi} \oint \frac{dz_2}{2i\pi} e^{-it_1z_1-it_2z_2}\\  \nonumber && \frac{ \prod_{i=1}^n (1+\frac{t_1}{z_1-r_i}) \prod_{j=1}^m (1-\frac{t_2}{z_2-\rho_j})}{\prod_{j=1}^m ( 1+\frac{t_1}{z_1-\rho_j}) \prod_{i=1}^n( 1-\frac{t_2}{ z_2-r_i})} [1+ \frac{t_1t_2}{(z_1-z_2)(z_1-z_2+t_1+t_2)}] \ea 
 in which the contour for $z_1$ encircles the $r$-poles and $z_2$ the $\rho$-poles ;
\ba &&U^{III} (t_1,t_2) = \frac{e^{it_1^2/2 - it_2^2/2}}{t_1t_2} \oint \frac{dz_1}{2i\pi} \oint \frac{dz_2}{2i\pi} e^{-it_1z_1-it_2z_2}\\  \nonumber && \frac{ \prod_{i=1}^n (1+\frac{t_2}{z_2-r_i}) \prod_{j=1}^m (1-\frac{t_1}{z_1-\rho_j})}{\prod_{j=1}^m ( 1+\frac{t_2}{z_2-\rho_j}) \prod_{i=1}^n( 1-\frac{t_1}{ z_1-r_i})} [1+ \frac{t_1t_2}{(z_1-z_2)(z_1-z_2-t_1-t_2)}] \ea  
$z_1$ encircles the $\rho$-poles and $z_2$ the $r$-poles,
 \ba &&U^{IV}(t_1,t_2) = \frac{e^{it_1^2/2 +it_2^2/2}}{t_1t_2}   \oint \frac{dz_1}{2i\pi}e^{-it_1z_1} \oint \frac{dz_2}{2i\pi}e^{-it_2z_2}\\ \nonumber && \frac{ \prod_{j=1}^m (1-\frac{t_1}{z_1-\rho_j}) (1-\frac{t_2}{z_2-\rho_j})}{ \prod_{i=1}^n ( 1-\frac{t_1}{z_1-r_i})( 1-\frac{t_2} {z_2-r_i})} [1+ \frac{t_1t_2} {(z_1-z_2-t_1) (z_1-z_+-t_2)}]\ea
 $z_1$ and $z_2$ encircle the $\rho$-poles. 
Remarkably enough these four terms recombine nicely into one single compact expression. First the ones which appear as first terms in the brackets reconstruct simply the disconnected part $U(t_1)U(t_2)$. Then after appropriate shifs $ z_i\to z_i \pm t_i/2$ the four integrands become identical and their sum is simply obtained by taking the residues at all the poles in the $z_1, z_2$ plane. The final expression for the connected correlation function  is  then
\ba U_c(t_1,t_2)  &=&  \oint \frac{dz_1}{2i\pi} \frac{dz_2}{2i\pi}e^{-it_1z_1-it_2z_2} \prod_1^n \frac {(z_1 -r_i +t_1/2)(z_2-r_i+t_2/2)}{(z_1 -r_i -t_1/2)(z_2-r_i-t_2/2)}\nonumber\\
&&\times \prod_1^m \frac{ (z_1 -\rho_j-t_1/2)(z_2-\rho_i-t_2/2)}{(z_1 -\rho_j +t_1/2)(z_2-\rho_j +t_2/2)}\nonumber\\
&&\times \frac {1}{ (z_1-z_2 -t_1/2-t_2/2)(z_1-z_2 +t_1/2+t_2/2)} \ea
It is clear that this may be generalized to a k-point function as in the usual GUE case \cite{BH1}.

 \vskip 2mm
 %%%%%%%%%%%%%%%%%%%%%%%%%%%%%%%%%
\section {Duality}
In the GUE case we have used at length a duality between  the expectation value of a product of  k-characteristic polynomials  with $N\times N$ random matrices in a source, which is equal to the expectation  values of the product of $N$ characteristic polynomials averaged with $k\times k$ random matrices \cite{BH1,BH2}. 
We now derive a similar duality for supermatrices. \\
Consider first  the one point expectation value
\be  F_1(x) = <\frac{1}{{\rm{sdet}}(x-M)}> = <\int d\Phi e^{i \bar \Phi (x-M) \Phi}> \ee 
with the weight  (\ref{PA}) ; the $(n+m)$-components  vector $\Phi$ consists of \\
$(u_1, \cdots, u_n; \theta_1,\cdots, \theta_m) $ with anticommuting $\theta$'s : $
d\Phi$ stands for \\
$\prod_idu^{\star}_i du_i\prod_j d\bar\theta_j d\theta_j$. 
The integral over the matrix $M$ with source $A$ is replaced by an integral with source 
\be \tilde A = A + \left(\begin{array}{cc}
u_iu{^\star}_j& u_i\bar \theta_j  \\ u_j^{\star} \theta_i& \theta_i \bar \theta_j 
\end{array}\right)\ee
Then
\be  F_1(x) =\int d\Phi e^{ix \bar \Phi \cdot \Phi} \frac {Z_{\tilde A}}{Z_A} = \int d\Phi e^{ix \bar \Phi \cdot \Phi} e^{-i/2 s\tr (\tilde A^2- A^2)} \ee 
and 
\be\frac{1}{2} s\tr (\tilde A^2- A^2) = \sum_1^n r_iu_iu^{\star} _i+\sum_1^m \rho_j \bar \theta_j\theta_j + \frac{1}{2}(u^{\star}\cdot u)^2 +  (u^{\star}\cdot u)(\bar \theta \cdot \theta) + \frac{1}{2}(\bar \theta \cdot \theta)^2 \ee
Using the representation
\be e^{-i/2 ( u^{\star}\cdot u + \bar \theta \cdot \theta)^2 } = \int dy e^{i y^2/2 +i y  (u^{\star}\cdot u + \bar \theta \cdot \theta)} \ee
(up to normalizations), we can now integrate out the $u's$ and $\theta 's$ and end up with a single integral
\be  F_1(x) = <\frac{1}{{\rm{sdet}}(x-M)}> = \int dy e^{-iy^2/2 } \prod_1^n \frac{1}{x+y-r_i} \prod_1^m (x+y -\rho_j) \ee
over the variable $y$ ; shifting $y\to y-x$ we end up with
\be \label{1det}  <\frac{1}{{\rm{sdet}}(x-M)}> = e^{ix^2/2} \int dy e^{iy^2/2  +ixy} \frac {\prod_{1}^m (y-\rho_j)}{\prod_{1}^n(y-r_i)} \ee
In this dual representation we could introduce a $2\times 2$ diagonal supermatrix with  non-zero elements $\prod_1^n(y-r_i)$ and 
$  \prod_1^m (y-\rho_j)$ on the diagonal  and the fraction in (\ref{1det}) replaced by $1/{\rm{sdet}}$ to make the duality more explicit.

The same technique may be applied to
\be F_k(x_1\cdots x_k) = < \prod_{a=1}^k {\rm{sdet}}(x_a-M)^{-1} > \ee
i.e. 
\be F_k(x_1\cdots x_k) = < \prod_{a=1}^k \int d\Phi_a e^{i \bar \Phi_a (x_a-M) \Phi_a} > \ee
We are now dealing with a modified matrix source

\be \tilde A = A + \sum_{a=1} ^k\left(\begin{array}{cc}
u^a_iu^a{^\star}_j& u^a_i\bar \theta^a_j  \\ u^{ a\star}_j \theta^a_i& \theta^a_i \bar \theta^a_j 
\end{array}\right)\ee
The result of the integration over the matrix $M$  produces again $e^{-i/2 s\tr (\tilde A^2- A^2)}$ which involve quartic terms in $u$'s and $\theta$'s. 
 The Gaussian disentanglement of those fourth order terms involves now a $k \times k$ matrix $y_{ab}$ and we end up with
\be F_k(x_1\cdots x_k)  = e^{i\sum_1^k x_a^2/2} \int d y_{ab} e^{i/2\tr y^2-i\sum_a x_ay_{aa}} \frac { \prod_1^m \det(y-\rho_j)}{\prod_1^n \det(x-r_i)} \ee
which we could again express as the superdeterminant of a $2k\times 2k$ supermatrix. .

%%%%%%%%%%%%%%%%%%%%%%%%%%%%%
\section{What can we learn from supermatrices?}
At this stage it is natural to ask whether the whole machinery which has been developped over the years with usual matrix models, such as triangulations of random surfaces, planar limit, multicritical points, double scaling limit, intersection numbers of curves on Riemann surface, etc,  lead to something new with supermatrices. For instance consider a matrix model with a weight
\be P(M) = \frac{1}{Z} e^{ s\tr V(M)} \ee
in which V is a polynomial with complex coefficients.
Integrating out the $U(n\vert m)$ degrees of freedom one has 
\be Z = \int \prod_1^n dl_i \prod_1^m d\mu_j  ( \frac {\Delta(l) \Delta(\mu)}{\Delta(l\vert\mu)})^2\ e^{\sum_iV(l_i) -\sum_jV(\mu_j)} \ee
Introducing the densities
\be \rho_1(\la) = \frac{1}{ n} \sum_1^n \delta(\la-l_i) \hskip1cm \rho_2(\mu) = \frac{1}{ m} \sum_1^m \delta(\mu-\mu_j)\ee
we obtain
\ba &&Z =  \int \prod_1^n dl_i \prod_1^m d\mu_j  e^{n \int d\la \rho_1(\la)V(\la) -m\int d\mu\rho_2(\mu)V(\mu)}  \nonumber \\&& e^{  n^2 \int d\la d\la' \rho_1(\la)\rho_1(\la') \log{\vert \la - \la' \vert} +m^2 \int d\mu d\mu' \rho_2(\mu)\rho_2(\mu') \log{\vert \mu - \mu' \vert} -2nm \int d\la d\mu \rho_1(\la)\rho_2(\mu) \log{\vert \la - \mu\vert }}\nonumber\\ \ea
So if we define
\be \tilde\rho (\la) = n\rho_1 (\la) -m\rho_2(\la) \ee
the integral for the partition function takes the same form as the usual matrix model with
\be \int D\tilde \rho(\la) e^{ \int d\la V(\la) \tilde \rho(\la) + \int d\la d\la'\tilde \rho(\la)\tilde \rho(\la') \log{ \vert \la-\la'\vert}}\ee
Therfeore it seems that there are no modifications with respect to the usual matrix model, at least in the planar limit : the $m$ Grassmanian dimensions have simply reduced the number of commuting dimensions to $(n-m)$.  

However the situation for the model with external source, which in the usual case was useful for computing  intersection numbers, is slightly different.

\vskip 2mm

 \section{Intersection numbers for $p$-spin curves}
 \vskip 2mm
 The ordinary intersection numbers of the moduli space of curves may be  derived from a  generalization of Kontsevich' Airy matrix model \cite{Kontsevich}.
 The intersection numbers for one marked point for $p$-spin curves are computed from $U(t)$ by an appropriate tuning of 
 the external source \cite{BH1}.  When $p=2$, we obtain simply the Kontsevich' Airy model.
 
 For  supermatrices  the one-point function $U(t)$ is given by  (\ref{one}). We shall now tune the external parameters $r_i$ and $\rho_j$ (i=1,...,n, j=1,...,m). Define the sum 
 \be
c_k = \sum_{i=1}^n \frac{1}{r_i^k} - \sum_{j=1}^m \frac{1}{\rho_j^k}
\ee
$k$ is an integer and 
 expand $U(t)$ of (\ref{one}) as, 
 \be\label{1pointp}
 U(t) = \frac{1}{z}\int \frac{dz}{2i\pi} [ e^{- \frac{c_{p+1}}{p+1} ((z+\frac{t}{2})^{p+1} - (z-\frac{t}{2})^{p+1})} ]
 \ee
 where we have  chosen the $r's$ and $\rho's$ satisfying the conditions 
 \be
 c_1 = 0, \hskip 3mm c_2 = - i
 \ee
 \be
 c_j= 0,  \hskip 3mm (j=3,...,p)
 \ee
 
 The higher terms proportional  to $c_{k}$ ($k> p+1$) can be dropped in an appropriate scaling region with $n$ and $m$ large. We assume $c_{p+1} \sim (n -m)$, which is large. In the case  $n=m$, it reduces to $p= - 1$, which is expicited below. The new term is the second term in (\ref{one}) compared with the
 ordinary $p$-spin intersection numbers \cite{BH1}.
 
 For $p=-1$, we find like in (\cite {BH1}) the Euler characteristics $\chi(M_{g,n})$ . From (\ref{1pointp}),
 \be
 U(t) =\frac{1}{t} \int \frac{dz}{2i\pi} (\frac{z-1}{z+1})^N 
 \ee
 where $N= c_{p+1}$. By the change of variable, $(z-1)/(z+1) = e^{-y}$, we compute 
 \be
 U(t) = \sum \frac{\tilde B_{n}}{n}(\frac{1}{N})^n (-1)^n 
 \ee
Denoting $\tilde B_1 = \frac{1}{2}$, and $\tilde B_{j}=0$ for j odd, (j $>$1),
 we obtain the same intersection numbers as for the ordinary case with an overall factor 2. ($\tilde B_{2n} = B_n (-1)^{n+1}$, and Bernoulli number $B_n= 2n \zeta(1-2n)(-1)^n$).
 Thus we have obtained the Euler characteristics for the one puncture  cas,  equal to what was  derived from the GUE matrix model with source \cite{BH1},
 \be
 \chi (M_{g,1}) = \zeta(1 - 2g)
 \ee
 
For $p=2$ and $q=-1$ case, we obtain a natural extension of the  Kontsevich-Penner model, related now to open intersection numbers. It is not necessary to deal with 
quantum mechanical matrix models,  or two matrix models,  as was done  in \cite{BH1}. This is an advantage of the supermatrices formulation.
\vskip 2mm
\
\section{Supermatrices UOSp(n$\vert$m) and open boundaries}
\vskip 2mm

In \cite{BH1, BH3} we had considered the non-orientable triangulated surfaces generated by matrix models with matrices drawn from  the  Lie algebras of  $O(N)$ and $Sp(N)$. For such algebras the HarishChandra formula \cite{HC} allowed us to repeat all the steps followed for the unitary model.  We had obtained explicitly the n-point function $U(t_1,...,t_n)$. Thereby , after tuning of the external source, this yields generating functions for  topological invariants such as the virtual Euler characteristics and the
 intersection numbers. For  non-orientable surfaces, one cannot introduce  the first Chern class  since the direction of the spin can not be defined. However, in our previous study \cite{BH2,BH3} based on these Lie algebras, we have found , in analogy with the unitary model,  generalizations of the  topological invariants.  It is thus natural to conjecture that they correspond to  intersection numbers for non-orientable Riemann surfaces.\\

It is interesting to generalize these non-orientable surfaces to super-surfaces generated by a matrix model based on the super-unitary orthosymplectic  Lie algebra $UOSp(n\vert m)$.
The extension can be easily done   with the modification of  the HarishChandra (Itzykson-Zuber)  formula for unitary supermatrices that we used in the above section 2.
 
 The random matrix $M$ belonging to  $UOSp(n\vert m)$ and the external source $A$ are diagonalized by  unitary orthosymplectic  matrices  $U$, $V$ $\in$ $UOSp(n\vert m)$ 
\be U^{\dagger}M U = \left(\begin{array}{cc}
l & 0 \\ 0 & \mu
\end{array}\right), \hskip 5mm  V^{\dagger}A V = \left(\begin{array}{cc}
r & 0 \\ 0 & \rho
\end{array}\right) \ee

The extension of the  HarishChandra formula to superLie algebras has been derived by Guhr \cite{Guhr},
\ba\label{IZ2}
 I &=& \int_{U\in UOSp(2n\vert m)} dU e^{is\tr
 U^{\dagger} D U A}\nonumber\\
 &=&\frac{ ( \det [{\rm cos}(2l_i r_j)] + {\rm det}[ i {\rm sin}(2 l_i r_j)]) \det [ - 2 i {\rm sin}( 2 \mu_i \rho_j )]   \Delta(l^2 \vert\mu^2) \Delta( r^2 \vert\rho^2)} {\Delta(l^2)\Delta(\mu^2)  \Delta(r^2)\Delta(\rho^2) \prod \mu_j\prod \rho_j }\nonumber\\
  \ea
up to a normalization.
After integrating out these "angular" degrees of freedom one obtain an integral over the eigenvalues $l$'s and $\mu$'s of the random  matrices with the new "Berezinian" 
\be J(l, \mu) = [\frac { \Delta(l^2) \Delta(\mu^2) \prod \mu_j}{ \Delta (l^2\vert \mu^2)} ]^2 \ee
with
\be  \Delta (l^2\vert \mu^2) = \prod_1^n \prod_1^m( l_a^2-\mu_b^2) \ee
Using the above formulae, one  obtains the one point  function $U(t) = < s\tr e^{itM} >$,
\ba \label{U} &&U(t) = \frac{\Delta(r^2 \vert \rho^2)}{Z_A\Delta(r^2)\Delta(\rho^2) \prod \rho_j} \int \prod dl_i d\mu_j ( \sum_a \cos (t l_a)- \sum_b \cos (t \mu_b))\nonumber\\
&&\frac{ \Delta(l^2) \Delta(\mu^2) \prod \mu_j}{\Delta(l^2 \vert \mu^2)} 
e^{{i}(\sum l_i^2 -  \sum \mu_j^2) + 2i\sum l_ir_i-2i\sum\mu_j\rho_j}
\ea
The $\cos{tl}$ and $\cos{t\mu}$ lead to a split 
\be U(t) =U^I(t) + U^{II}(t) \ee
\be \label{1pointosp}U^I(t) = \frac{e^{-it^2/4}}{t}  \oint \frac{dz}{2i\pi}e^{-itz} \prod_{i=1}^n \frac{(z+\frac{t}{2})^2 -r_i^2}{z^2-r_i^2}
\prod_{j=1}^m \frac{z^2 -\rho_j^2}{(z+\frac{t}{2})^2-\rho_j^2} (\frac{z}{z + \frac{t}{4}})
\ee
where the contour encircles all  the poles at $z=\pm r_i$. This expression is similar to the one that was derived  with the  $O(N)$ antisymmetric real matrices  \cite{BH1}.
The second term gives 
\be  U^{II}(t) = \frac{e^{it^2/4}}{t}  \oint \frac{dz}{2i\pi}e^{-itz} \prod_{i=1}^m \frac{(z-\frac{t}{2})^2 -\rho_i^2}{z^2-\rho_i^2}
\prod_{j=1}^n \frac{r_i^2 -z^2}{r_i^2-(z-\frac{t}{2})^2}(\frac{z-\frac{t}{2}}{z-\frac{t}{4}}) \ee
where the contour is taken around all the poles at f z= $\pm \rho_j$.\\
This second term may  in fact be obtained from the first one for  $U^I(t)$  in (\ref{1pointosp}),  if  the contour in the z-plane  is  extended to encompass also the poles  at  $z= -\frac{t}{2} \pm \rho_j$ . 
Therefore, we obtain the  sum of the two terms $U(t) =U^I(t) + U^{II}(t)$   as a single contour integral.   After the shift $z\to z-\frac{t}{4}$ in (\ref{1pointosp}), the contour encircles now all the poles at $z=-\frac{t}{4}\pm r_i, \frac{t}{4}\pm \rho_j$, and
\be\label{OSp}
U(t)  =  \oint \frac{dz}{2i\pi}e^{-itz}  \prod_{i=1}^n (\frac{(z+\frac{t}{4})^2 -r_i^2}{(z-\frac{t}{4})^2-r_i^2}) 
\prod_{j=1}^m \frac{(z-\frac{t}{4})^2 -\rho_j^2}{(z+\frac{t}{4})^2-\rho_j^2} ( 1 - \frac{t}{4z})
\ee
One verifies that in the case  m=0,  this coincides withi the one point function of $O(N)$ case, and for  n=0, we obtain the  $Sp(N)$ result \cite{BH1}. The generalization to the $k$-point functions may easily follow  as was done hereabove in the unitary supersymmetric case.

A number of studies may be performed on the basis of these general formulae. We intend to consider the interesting case of  the $UOSp$ generalized Kontsevich model with a logarithmic term (open-boundary). This might be related to the geometry of super Riemann surfaces
with open boundaries,  but we leave the question to a subsequent work.

%%%%%%%%%%%%%%%%%%%%%%%%

\section{Summary}
We have investigated the k-point  correlation functions for the vertices str$e^{it M}$,  in a Gaussian ensemble invariant under $U(n\vert m)$. The  formulae that we have derived extend  the usual Hermitian  matrices results, with the freedom of two kind of external sources $r_i$ and $\rho_j$, bosonic and fermionic. This freedom allows one to compute  various topological invariants of surfaces,
for example, the intersection numbers with  boundaries, through an extension of the   Kontsevich-Penner model.
The extension to supermatrices $UOSp(n\vert m)$ is a generalization to  non-orientable surfaces generated by matrix models  based on the $O(N)$ or $Sp(N)$ Lie algebras.
The Kontsevich-Penner model obtained  from  supermatrices in  the Lie algebra of  $UOSp(n\vert m)$ may give more informations on manifolds with open boundaries, but this remains to be investigated  in a future work.
\vskip 2mm
{\bf Acknowledgement}
\vskip 2mm
\noindent We thank Edward Witten for  communication of his  manuscript \cite{DijkgraafWitten} and  for formulating interesting questions, which led us to investigate supermatrices. S.H. thanks the  support by JSPS
KAKENHI (C) 16K05491.

\end{document}